\documentclass[twocolumn,aps,prl,superscriptaddress,nofootinbib]{revtex4-2}
\usepackage{graphicx}
\usepackage{subcaption}
\usepackage{color}
\graphicspath{{figures/}{fig/}}
\usepackage{amsmath}
\usepackage{amssymb}
\usepackage{bm}
\usepackage{slashed}
\usepackage{epsfig}
\usepackage{amsfonts}
\usepackage{epstopdf}
\usepackage{hyperref}
\usepackage{bbm}
\usepackage{textcomp}
\usepackage{color}
\usepackage{array}

\newcommand{\sect}[1]{{\it \textbf{#1.} --- }}
\newcommand{\mi}{\mathrm{i}}

\begin{document}
\title{ Tame multi-leg Feynman integrals beyond one loop}

\author{Li-Hong Huang}
\email{lhhuang@pku.edu.cn}
\affiliation{School of Physics, Peking University, Beijing 100871, China}

\author{Rui-Jun Huang}
\email{huangrj2215@pku.edu.cn}
\affiliation{School of Physics, Peking University, Beijing 100871, China}

\author{Yan-Qing Ma}
\email{yqma@pku.edu.cn}
\affiliation{School of Physics, Peking University, Beijing 100871, China}
\affiliation{Center for High Energy Physics, Peking University, Beijing 100871, China}

\date{\today}

\begin{abstract}
We introduce a novel structure for Feynman integrals, reformulating them as integrals over a small set of parameters with a fully controllable integrand. The integrand closely resembles one-loop Feynman integrals, and they are very easy to handle. Remarkably, the number of remaining integration parameters is independent of the number of external legs and small—at most 2 for two-loop integrals and 5 for three-loop integrals—facilitating the application of a wide range of established methods, both specific to Feynman integrals and more general techniques. This approach is expected to mitigate the computational challenges of multi-loop, multi-leg Feynman integrals. As a proof of concept, we successfully computed two-loop non-planar Feynman integrals with six external legs, demonstrating high efficiency.
\end{abstract}

\maketitle
\allowdisplaybreaks

\sect{Introduction}
With upcoming collider experiments, such as the High-Luminosity Large Hadron Collider (HL-LHC), experimental precision for numerous physical processes will soon surpass the corresponding theoretical precision in particle physics \cite{Azzi:2019yne,Cepeda:2019klc}. This underscores the urgent need for improvements in theoretical precision. Among the pivotal challenges in this regard is the computation of Feynman integrals (FIs).

Typically, the computation of one-loop FIs is regarded as a resolved matter and can be effected for any given number of external legs \cite{tHooft:1978jhc, Passarino:1978jh, Bern:1993kr,vanOldenborgh:1989wn, Hahn:1998yk, Ellis:2007qk, vanHameren:2010cp, Denner:2016kdg, Cullen:2011kv, Patel:2015tea}. However, as we ascend to higher loop levels, the complexity experiences a staggering escalation. For example, at the two-loop level, computing FIs with more than four external legs is already a state-of-the-art challenge, while at the three-loop level, the frontier is integrals with more than three external legs.

The current strategy involves reducing FIs to linear combinations of a finite set of bases using such integration-by-parts identities~\cite{Chetyrkin:1981qh,Tkachov:1981wb,Laporta:2000dsw,Anastasiou:2004vj,Smirnov:2008iw,Smirnov:2023yhb,Studerus:2009ye,Lee:2013mka,Maierhofer:2017gsa, Peraro:2019svx, Klappert:2020nbg,Wu:2023upw,Guan:2024byi}, and then computing these bases \cite{Binoth:2000ps,Heinrich:2008si,Smirnov:1999gc,Kotikov:1990kg,Bern:1992em,Remiddi:1997ny,Gehrmann:1999as,Henn:2013pwa,Lee:2014ioa,Adams:2017tga,Liu:2017jxz}. Despite significant progress in recent years, the computation of multi-leg FIs beyond one-loop order persists as a major challenge. The primary difficulty lies in the fact that these integrals, regardless of whether they are expressed in the Feynman parameter or Baikov representation \cite{Baikov:1996iu}, involve a large number of integration parameters, rendering the reduction step extraordinarily intricate. For instance, the integration-by-parts reduction leads to a huge system of equations, while the reduction using asymptotic expansions  \cite{Baikov:2005nv,Liu:2018dmc} results in an excessive number of terms. This difficulty proves to be formidably to surmount without a more profound exploration and understanding of the underlying structure of FIs.

In this Letter, we disclose a novel structure of FIs. We find FIs can be represented as integrals over a significantly  diminished  number of parameters, while the integrand exhibits a simplicity akin to one-loop FIs. This newly unearthed structure offers powerful methodologies for computing FIs, thereby facilitating the fulfillment of the high-precision demands of theoretical predictions.

\sect{A new representation}
A $L$-loop FI, or amplitude, can be expressed as
\begin{align}
    \mathcal{M} \equiv \int \prod_{i=1}^{L} \frac{\mathrm{d}^D l_i}{\mathrm{i} \pi^{D/2}} \frac{P(l)}{\mathcal{D}_1^{\nu_1} \cdots \mathcal{D}_N^{\nu_N}},
\end{align}
where $D$ is the spacetime dimension, $\l_i$ are the loop momenta, and $P(l)$ is a numerator that depends polynomially on the loop momenta. The components of $\vec{\nu}$ are positive integers, and the denominators are given by
\[
\mathcal{D}_\alpha = \sum_{i,j=1}^{L} \mathcal{\hat A}^\alpha_{ij} \, l_i \cdot l_j + 2\sum_{i=1}^{L} \mathcal{\hat B}^\alpha_i \cdot l_i + \mathcal{\hat C}^\alpha,
\]
where a Feynman prescription $\mathrm{i} 0^+$ is suppressed.
Two propagators, $\mathcal{D}_\alpha $ and $\mathcal{D}_\beta$, are said to belong to the same branch if the coefficients $\mathcal{\hat A}^\alpha_{i,j}$ and $\mathcal{\hat A}^\beta_{i,j}$ are identical. We use $n_b$ to denote the number of denominators in the $b$-th branch.  The total number of branches, denoted by $B$, satisfies $B \leq 3L - 3$ when $L\geq2$ for ordinary Feynman integrals. For convenience, we assume that the propagators are ordered so that $\mathcal{D}_{n_1+\cdots+n_{b-1}+i}$ is the $i$-th propagator in the $b$-th branch, also denoted as $\mathcal{D}_{(b,i)}$.

By first introducing Feynman parameters $y_{(b,i)}$ to combine the denominators within the same branch, and then introducing Feynman parameters $X_b$ to combine all denominators, we obtain the following expression:
\begin{equation}
\begin{aligned}
    &\frac{1}{\mathcal{D}_1^{\nu_1} \cdots \mathcal{D}_N^{\nu_N}}\equiv\prod_{b=1}^{B}\prod_{i=1}^{n_b}\frac{1}{\mathcal{D}_{(b,i)}^{\nu_{(b,i)}}}=\frac{\Gamma(\nu)}{\prod_{\alpha=1}^{N}\Gamma\left(\nu_{\alpha}\right)} \\    &\times \int_0^\infty\left[\mathrm{d}\mathbf{X}\right]\left[\mathrm{d}\mathbf{y}\right]\frac{\prod_{b=1}^{B}X_b^{\nu_b-1} \prod_{\alpha=1}^{N} y_{\alpha}^{\nu_{\alpha}-1}}{\left(\sum_{b=1}^{B} \sum_{i=1}^{n_b} X_b y_{(b,i)}{\mathcal{D}}_{(b,i)}\right)^\nu},
\end{aligned}
\end{equation}
where $\nu_b=\sum_{i=1}^{n_b} \nu_{(b,i)}$,  $\nu=\sum_{\alpha=1}^{N} \nu_{\alpha}$, and the integration measures are defined as
\begin{align}    &\left[\mathrm{d}\mathbf{X}\right]=\prod_{b=1}^{B}\mathrm{d}X_b\delta\left(1-\sum_{b=1}^{B} X_b\right),
\\    &\left[\mathrm{d}\mathbf{y}\right]\equiv \prod_{\alpha=1}^{N}\mathrm{d}y_{\alpha}\prod_{b=1}^{B}\delta\left(1-\sum_{i=1}^{n_b} y_{(b,i)}\right).
\label{eq:dy}
\end{align}
The combination of denominators can be expanded as
\begin{equation}\label{eq:sumxD}
    \sum_{b=1}^{B} \sum_{i=1}^{n_b} X_b y_{(b,i)}{\mathcal{D}}_{(b,i)}=\sum_{i,j=1}^{L}\mathcal{A}_{ij} \; l_i\cdot l_j+2\sum_{i=1}^{L}\mathcal{B}_i\cdot l_i+\mathcal{C}\,.
\end{equation}
Naively, both $\mathcal{A}_{ij}$, $\mathcal{B}_i$ and $\mathcal{C}$ should be bilinearly dependent on $\mathbf{X}$ and $\mathbf{y}$. However, due the delta functions in Eq.~\eqref{eq:dy}, the symmetric matrix $\mathcal{A}$ is independent of $\mathbf{y}$.

It is convenient to introduce the Symanzik polynomials $\mathcal{U}$ and $\mathcal{F}$ defined as
\begin{equation}
    \begin{aligned}
        &\mathcal{U}=\det\left(\mathcal{A}\right),\\
        &\mathcal{F}=\left(\mathcal{B}_\mu\right)^T \mathcal{A}^{adj}\mathcal{B}^\mu-\mathcal{C}\det\left(\mathcal{A}\right),
    \end{aligned}
\end{equation}
where $\mathcal{A}^{adj}$ denotes the adjugate matrix of $\mathcal{A}$. From the discussion above, $\mathcal{U}$ is a homogeneous polynomial of degree $L$ in $\mathbf{X}$ and is independent of $\mathbf{y}$. By performing the substitution in $\mathcal{C}$ with $y_{(b,i)}\rightarrow y_{(b,i)}\times 1=y_{(b,i)}\sum_j y_{(b,j)}$, $\mathcal{F}$ becomes a homogeneous polynomial of degree 2 in $\mathbf{y}$, which can be  written as
\begin{equation}
    \mathcal{F}=\frac{1}{2}\sum_{\alpha,\beta=1}^N R_{\alpha\beta}\;y_\alpha y_{\beta}=\frac{1}{2} \mathbf{y}^T \cdot  R   \cdot  \mathbf{y},
    \label{eq:F2R}
\end{equation}
where  $R$ is a symmetric matrix with elements being homogeneous polynomials of degree $L+1$ in $\mathbf{X}$.

By making the substitution $l_i^{\mu}\to l_i^{\mu}-\left(\mathcal{A}^{-1}\cdot \mathcal{B}^{\mu} \right)_i$ to eliminate the linear terms in $l$ in Eq.~\eqref{eq:sumxD}, we eventually obtain a new representation of FIs:
\begin{equation}    
\mathcal{M}=\int\left[\mathrm{d}\mathbf{X}\right] \hat{\mathcal{M}}\left(\mathbf{X}\right),
\label{eq:rep}
\end{equation}
where the integrand is defined as
\begin{equation}
    \begin{aligned}
    \hat{\mathcal{M}}&\left(\mathbf{X}\right)=\frac{\Gamma(\nu)}{\prod_{\alpha=1}^{N}\Gamma\left(\nu_{\alpha}\right)} \prod_{b=1}^{B} X_b^{\nu_b-1}\int\left[\mathrm{d}\mathbf{y}\right]\prod_{\alpha=1}^{N} y_{\alpha}^{\nu_{\alpha}-1}\\
    &\times  \int \prod_{i=1}^{L} \frac{\mathrm{d}^D l_i}{\mathrm{i} \pi^{D/2}} \frac{P(l^{\mu}-\mathcal{A}^{-1}\cdot \mathcal{B}^{\mu})}{(\sum_{i,j=1}^{L}\mathcal{A}_{ij} \; l_i\cdot l_j-\mathcal{F}/\mathcal{U})^\nu}.
\end{aligned}\label{eq:Jtilde}
\end{equation}
After performing the straightforward integration over the loop momenta, $\hat{\mathcal{M}}$ can be expressed as
\begin{align}
    \hat{\mathcal{M}}\left(\mathbf{X}\right)=\mathcal{U}^{-\frac{(L+1)D}{2}}\sum_{\Delta, \vec{\nu}^\prime} K^{\Delta}_{\vec{\nu}^\prime}(\mathbf{X})\, I_{\Vec{\nu}^\prime}^{\Delta}(\mathbf{X}),
\end{align}
where $\Delta$ are numbers related to spacetime dimension, the simple coefficients $K$ depend rationally on $\mathbf{X}$ and the kinematic variables, and the $I$ are  defined as
\begin{equation}
I_{\Vec{\nu}}^{\Delta}(\mathbf{X})=\frac{(-1)^\nu\Gamma(\nu-\Delta)}{\prod_{\alpha=1}^{N}\Gamma\left(\nu_{\alpha}\right)}\int\left[\mathrm{d}\mathbf{y}\right]\frac{\prod_{\alpha=1}^{N}y_{\alpha}^{\nu_{\alpha}-1}}{\left(\frac{1}{2} \mathbf{y}^T\hspace{-1mm}\cdot\hspace{-1mm} R \hspace{-0.5mm} \cdot \hspace{-0.5mm} \mathbf{y} -\mathrm{i}0^+ \right)^{\nu-\Delta}},
\label{eq:FBI}
\end{equation}
which are integrals with branch parameters $\mathbf{X}$ fixed, referred to as \textit{fixed-branch integrals} (FBIs).

Since FBIs are structurally similar to one-loop integrals in Feynman parameter representation, they can be easily handled, as will be demonstrated later. Thus, the new representation in Eq.~\eqref{eq:rep} highlights an important feature: any multi-loop Feynman integral, or amplitude, can be reformulated as an integral over a small number of branch parameters, with the integrand fully under control. Since the number of branch parameters is $B-1$, accounting for a delta function, and is independent of the number of external legs, this new representation offers much more powerful methods for evaluating multi-leg Feynman integrals beyond one loop.

\sect{Evaluation of FBIs}
As $\mathbf{X}$ are fixed parameters in FBIs, we will suppress their dependence  during the computation of FBIs.
Like Feynman integrals, FBIs can be reduced to a small set of basis integrals, referred to as the master integrals (MIs) of FBIs. 

Following and generalizing the one-loop notation in Ref.~\cite{Duplancic:2003tv}, we introduce a symmetric matrix $S$ defined as follows: 1) $S_{\alpha,\beta}=R_{\alpha-B,  \beta-B}$ if $\alpha>B$ and $\beta>B$; 2) $S_{\alpha,\beta}=1$ if $\alpha\leq B$ and $1\leq\beta-B-\sum_{b = 1}^{\alpha - 1}n_b\leq n_\alpha$, or $\beta\leq B$ and $1\leq\alpha-B-\sum_{b = 1}^{\beta - 1}n_b\leq n_\beta$; 3) $S_{\alpha,\beta}=0$ otherwise. For instance, if $B=3$ and $(n_1,n_2,n_3)=(2,1,1)$, the matrix S corresponding to this integral is
\begin{equation}
    S=\begin{pmatrix}
        0_{3\times3}&\begin{matrix}
                     1&1&0&0\\
                    0&0&1&0\\
                    0&0&0&1
                \end{matrix}\\
        \begin{matrix}
            1&0&0\\
            1&0&0\\
            0&1&0\\
            0&0&1
        \end{matrix}&R
    \end{pmatrix}.
    \label{eq:S}
\end{equation}

Using integration by parts, we obtain two relations (derivation provided in the Supplementary Material). The first one is a recursion relation
\begin{equation}
\begin{aligned}
    S \cdot&(t_1,\cdots,t_B,\nu_1 I_{\vec{\nu}+\vec{e}_{1}}^{\Delta},\cdots,\nu_N I_{\vec{\nu}+\vec{e}_{N}}^{\Delta})^T\\
    =&(-I_{\vec{\nu}}^{\Delta-1},\cdots,-I_{\vec{\nu}}^{\Delta-1},I_{\vec{\nu}-\vec{e}_{1}}^{\Delta-1},\cdots,I_{\vec{\nu}-\vec{e}_{N}}^{\Delta-1})^T,
\end{aligned}\label{eq:rec}
\end{equation}
where $t_b$ as free parameters determined by the equation itself. And the second one is a dimension-shift relation: 
\begin{align}
    C I_{\vec{\nu}}^{\Delta-1}=\left(2\Delta-\nu-B\right) z_0 I_{\vec{\nu}}^{\Delta}+\sum_{\alpha=1}^{N} z_\alpha I_{\vec{\nu}-\vec{e}_\alpha}^{\Delta-1},
    \label{eq:dshift}
\end{align}
with $C=\sum_{b=1}^B C_b$ and the unknowns are constrained by
\begin{equation}\label{eq:Cz}
    S \cdot(C_1,\cdots,C_B,z_1,\cdots,z_N)^T=(z_0,\cdots,z_0,0,\cdots,0)^T\,.
\end{equation}
If $\det(S)\neq0$, we set $z_0=1$ and then the constants $C_b$ and $z_\alpha$ are fully determined. If $\det(S)=0$, we set $z_0=0$ and then $C_b$ and $z_\alpha$ belong to any basis of the null space of $S$, from which we choose a basis such that $C$ is as nonzero as possible.

With the above choices, we classify the given sector into four types:
\begin{enumerate}
\item $\det(S)\neq 0$ and $C\neq 0$:  In this case we use Eq.~\eqref{eq:rec} to reduce all FBIs to corner FBI, defined by $\nu_i=1$ for all $i$, as well as to FBIs in subsectors, defined by $\nu_i\leq0$ for some $i$. Additionally, Eq.~\eqref{eq:dshift} can be used to shift integrals to the same dimension.  Thus there is exactly one MI in this sector. 
\item $\det(S)\neq 0$ and $C= 0$: We can still use Eq.~\eqref{eq:rec} to reduce all FBIs to corner FBI. However, the relation Eq.~\eqref{eq:dshift} becomes
\begin{equation}
    \left(2\Delta-\nu-B\right) I_{\vec{\nu}}^{\Delta}=-\sum_{\alpha=1}^{N} z_\alpha I_{\vec{\nu}-\vec{e}_\alpha}^{\Delta-1},
\end{equation}
which reduces the value of $\nu$ for any FBI in this sector, including the corner FBI. Therefore, there is no MI in this sector.
\item $\det(S)= 0$ and $C\neq 0$: In this case Eq.~\eqref{eq:dshift} becomes
\begin{equation}
    C I_{\vec{\nu}}^{\Delta-1}=\sum_{\alpha=1}^{N} z_\alpha I_{\vec{\nu}-\vec{e}_\alpha}^{\Delta-1},
    \label{eq:ibptype3}
\end{equation}
which also always reduce the value of $\nu$ in this sector, resulting no MI in this sector.
\item $\det(S)= 0$ and $C= 0$: In this case at least one of $z_\alpha$ terms is nonzero.
Suppose that $|z_\beta|$ is the largest among these values, then Eq.~\eqref{eq:dshift} becomes
\begin{equation}
    I_{\vec{\nu}}^{\Delta}=-\sum_{\alpha\neq \beta}  \frac{z_\alpha}{z_\beta} I_{\vec{\nu}+\vec{e}_\beta-\vec{e}_\alpha}^{\Delta},
\end{equation}
which decreases the values of other $\nu_\alpha$ at the cost of increasing $\nu_\beta$. Ultimately, this can reduce all FBIs to subsectors, leaving no MI in this sector.
\end{enumerate}
Therefore, there is at most one MI in each sector, and the relations in Eqs. \eqref{eq:rec} and \eqref{eq:dshift}  are sufficient to efficiently reduce all FBIs to their MIs. Note that the reduction relations can also setup close differential equations of MIs with respect to $\mathbf{X}$ and kinematic variables, because derivative of these MIs are also linear combinations of FBIs.

To compute the MIs of FBIs, we apply the auxiliary-mass-flow method originally proposed in Ref.~\cite{Liu:2017jxz}.
Let us define a general FBI with an auxiliary parameter $\eta$:
\begin{equation}
\mathcal{I}_{\Vec{\nu}}^{\Delta}(\eta)=\frac{(-1)^\nu\Gamma(\nu-\Delta)}{\prod_{\alpha=1}^{N}\Gamma\left(\nu_{\alpha}\right)}\int\left[\mathrm{d}\mathbf{y}\right]\frac{\prod_{\alpha=1}^{N}y_{\alpha}^{\nu_{\alpha}-1}}{\left(\frac{1}{2} \mathbf{y}^T\hspace{-1mm}\cdot\hspace{-1mm} R \hspace{-0.5mm} \cdot \hspace{-0.5mm} \mathbf{y} +\eta\right)^{\nu-\Delta}},
\label{eq:FBI}
\end{equation}
which can be obtained from $I_{\Vec{\nu}}^{\Delta}$ by replacing $R_{\alpha\beta}\to R_{\alpha\beta}+2\eta/B^2$.
Then the right-hand-side of Eq.~\eqref{eq:Cz} will be changed to 
$(z_0,\cdots,z_0,-2\eta z_0/B,\cdots,-2\eta z_0/B)^T$ if we keep $S$ unchanged, and
the dimension-shift relation for $\mathcal{I}_{\Vec{\nu}}^{\Delta}(\eta)$ is slightly modified from Eq.~\eqref{eq:dshift}, by replacing $C$ by $C-2z_0 \eta$, if we keep $C$ unchanged.
Since $\frac{\mathrm{d}}{\mathrm{d}\eta}\mathcal{I}_{\vec{\nu}}^{\Delta}(\eta)=-\mathcal{I}_{\vec{\nu}}^{\Delta-1}(\eta)$, the dimension-shift relation leads to the following differential equation
\begin{equation}
    \begin{aligned}
        &(2z_0 \eta-C)\frac{\mathrm{d}}{\mathrm{d}\eta}\mathcal{I}_{\vec{\nu}}^{\Delta}(\eta)\\
        = &\left(2\Delta-\nu-B\right)z_0 \mathcal{I}_{\vec{\nu}}^{\Delta}(\eta)+\sum_{\alpha=1}^{N} z_\alpha \mathcal{I}_{\vec{\nu}-\vec{e}_\alpha}^{\Delta-1}(\eta) .
    \end{aligned}\label{eq:etaDE}
\end{equation}
Since $\mathbf{y}$ are finite, the value of $\mathcal{I}_{\vec{\nu}}^{\Delta}(\eta)$ as $\eta\to \infty$ can be easily achieved based on its definition in Eq.~\eqref{eq:Jtilde}. By solving the differential equation with the boundary condition at $\eta\to \infty$, we can obtain $\mathcal{I}_{\vec{\nu}}^{\Delta}(\eta)$ at any value of $\eta$. Finally, we have a dimension-changing transform \cite{DCT}
\begin{equation}
\begin{aligned}
     I_{\vec{\nu}}^{\Delta+\delta}&=\frac{1}{\Gamma(\delta)}\int_{-\mathrm{i}0^+}^{-\mathrm{i}\infty}\mathrm{d}\eta \;\eta^{\delta-1}\mathcal{I}_{\vec{\nu}}^{\Delta}(\eta),
\end{aligned}
\end{equation}
which allows us to compute the desired FBI at any spacetime dimension defined by $\Delta+\delta$, starting from $\mathcal{I}_{\vec{\nu}}^{\Delta}(\eta)$ in a specific spacetime dimension defined by $\Delta$.

It is important to emphasize that one-loop Feynman integrals in the Feynman parameter representation are a special case of FBIs with $B=1$. Furthermore, the value of $B$ is only an unimportant parameter in the above strategy for computing FBIs, we conclude that FBIs are as straightforward to compute as one-loop Feynman integrals.

\sect{Integration over branch variables}
Since FBIs can be easily computed, obtaining Feynman integrals only requires integrating out the branch variables in the representation given in Eq.~\eqref{eq:rep}. There are various approaches to achieve this, but here we explore only one method, leaving other alternatives for future investigation.

We begin by partitioning the integration into 
$B!$ regions, ensuring that $\mathbf{X}$ maintains a definite ordering within each region. 
For a region where $X_{i_1}<\cdots<X_{i_{B-1}}<X_{i_B}$, we perform the following substitution: 
$$X_1^\prime=X_{i_1} X_2^\prime,\cdots, X_{B-1}^\prime=X_{i_{B-1}}X_B^\prime, X_B^\prime=X_{i_B}.$$ 
After integrating out $X_B^\prime$ using the delta function,
the remained integration is over $\prod_{b=1}^{B-1}\mathrm{d}X_b^\prime$. To make our notation simpler, in the following replace $X_b^\prime$ by $X_b$, and express Eq.~\eqref{eq:rep} as 
\begin{equation}    
\mathcal{M}=\int_0^1\prod_{b=1}^{B-1}\mathrm{d}X_b \, \hat{\mathcal{M}}^\prime\left(\mathbf{X}\right),
\label{eq:Mhat}
\end{equation}
where $\hat{\mathcal{M}}^\prime\left(\mathbf{X}\right)$ denotes the sum of integrand over all $B!$ regions. 

The integral in Eq. \eqref{eq:Mhat} is divergent only on the surface defined by $X_1 X_2 \cdots X_{B-1}=0$, but the integrand may be singular at various surfaces, complicating numerical integration.
Before addressing these singularities, we first identify them.
Based on the definition in Eq.~\eqref{eq:FBI}, it is evident that $\eta$ can regularize all singularities in FBIs. Thus, the differential equation with respect to $\eta$ in Eq.~\eqref{eq:etaDE} contains all singular surfaces which have been regularized by $\eta$. Since $z_0=0$ when $\det(S)=0$, all singularities are determined by $C+\mi 0^+$ in sectors with $\det(S)\neq0$. We collect these $C$'s and factorize them to find all nontrivial polynomial factors, denoted as $P_j+\mi 0^+$.  Consequently, all singularities of the FBIs under consideration are generated by these $P_j$'s.

As usual, we deform the integration contour to eliminate singularities that are not on the boundary. We choose the deformation as
\begin{equation}
    \begin{aligned}
        &\tilde{X}_b=X_b+\mathrm{i}X_b(1-X_b)G_b(\mathbf{X}), 
    \label{eq:deform}
    \end{aligned}
\end{equation}
with $b=1, \cdots, B-1$, and
\begin{equation}
    \begin{aligned}
        &G_b(\mathbf{X})=\kappa \sum_j \lambda k_j\frac{\mathrm{\partial}_{X_b} P_j}{P_j^2+(\mathrm{\partial}_{X_b} P_j)^2}\exp({-\frac{P_j^2}{\lambda^2 k_j^2}}).
    \end{aligned}
    \label{eq:Gb}
\end{equation}
One property of this choice is that the deformation induced by a polynomial $P_j$ can only affect regions near $P_j=0$. This ensures that the derivatives of two polynomials, say $\mathrm{\partial}_{X_b}P_i$ and $\mathrm{\partial}_{X_b}P_j$, which may even have different signs, can always individually be deformed in the correct direction. Note that at the surface where $P_i=P_j=0$, both are for sure deformed in the same direction. Another advantage of this choice is that the magnitude of the deformation is of the same order as the affected size. To compute MIs of FBIs after contour deformation, we setup differential equations of MIs with respect to $\kappa$, expressed by differential equations of MIs with respect to $\mathbf{X}$ derived from reduction, and use $\kappa=0$ as boundary condition, which are MIs without deformation. Eventually, we choose $\kappa=1$.

To fix the remain free parameter in Eq.~\eqref{eq:Gb}, we begin by setting $\lambda = 10^{-2}$ and assigning all $k_j$ values to be sufficiently small, e.g., $10^{-1}$. For each $k_j$, we then determine its maximum allowed value, ensuring that the deformed contour does not intersect any singularity, while keeping the other parameters fixed at their initial values. Once we have the maximum values for all $k_j$, we update them accordingly and set $\lambda$ to a smaller value, e.g., $10^{-3}$. Finally, we adjust $\lambda$ to be half of its maximum allowed value.

After the contour deformation, Eq.~\eqref{eq:Mhat} can be rewritten as
\begin{equation}   \label{eq:intMtilde} 
\mathcal{M}=\int_0^1\prod_{b=1}^{B-1}\mathrm{d}X_b \, \tilde{\mathcal{M}}\left(\mathbf{X}\right).
\end{equation}
The singularities at boundary $X_b=0$ can be subtracted out taking advantage of differential equations with respect to $X_b$. Finally, the finite integration can be obtained by using numerical integration techniques.

\begin{table}[htb]
\centering
\begin{tabular}{| c | c |c |} 
\hline
graphs  & $\#$ points & DCT time/points (ms)  \\
\hline
\includegraphics[width=0.1\textwidth]{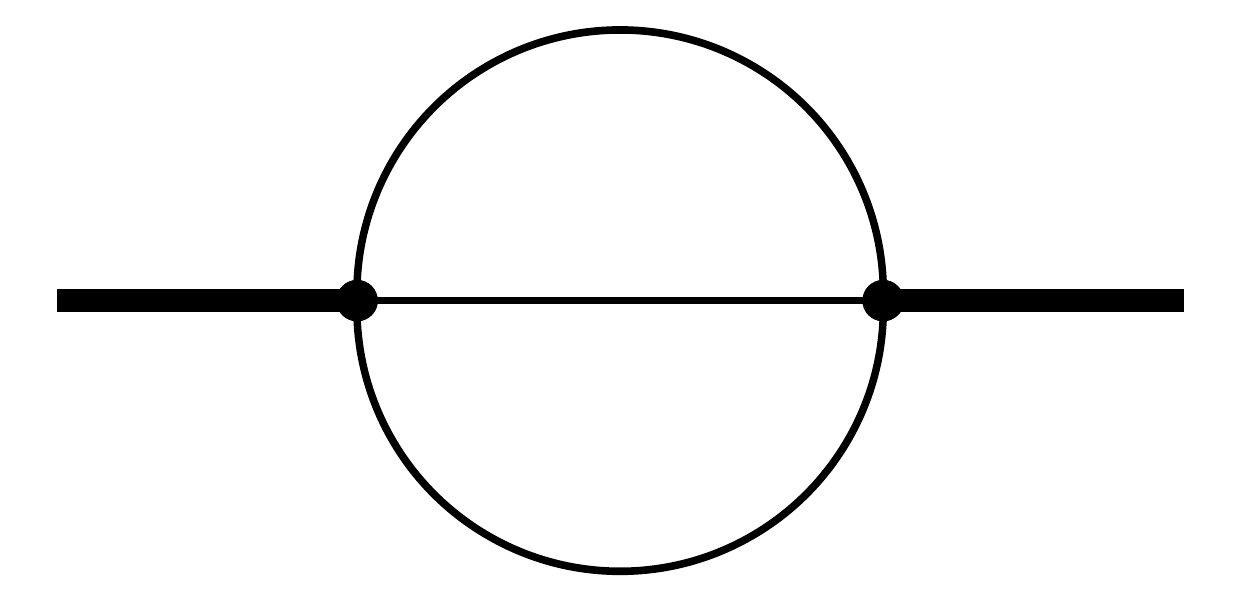} & \raisebox{1.5\height}{726} & \raisebox{1.5\height}{0.14}    \\
\hline
\includegraphics[width=0.1\textwidth]{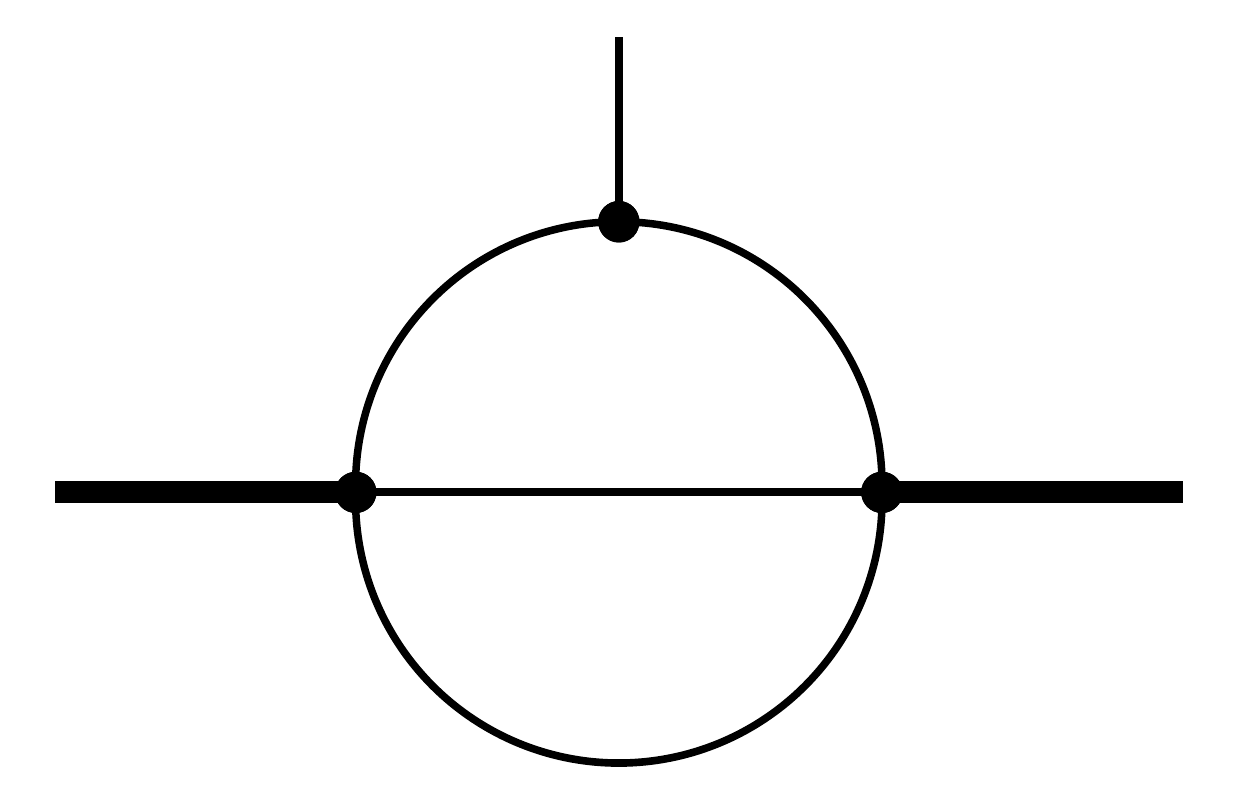} & \raisebox{2\height}{726} & \raisebox{2\height}{0.19}    \\
\hline
\includegraphics[width=0.1\textwidth]{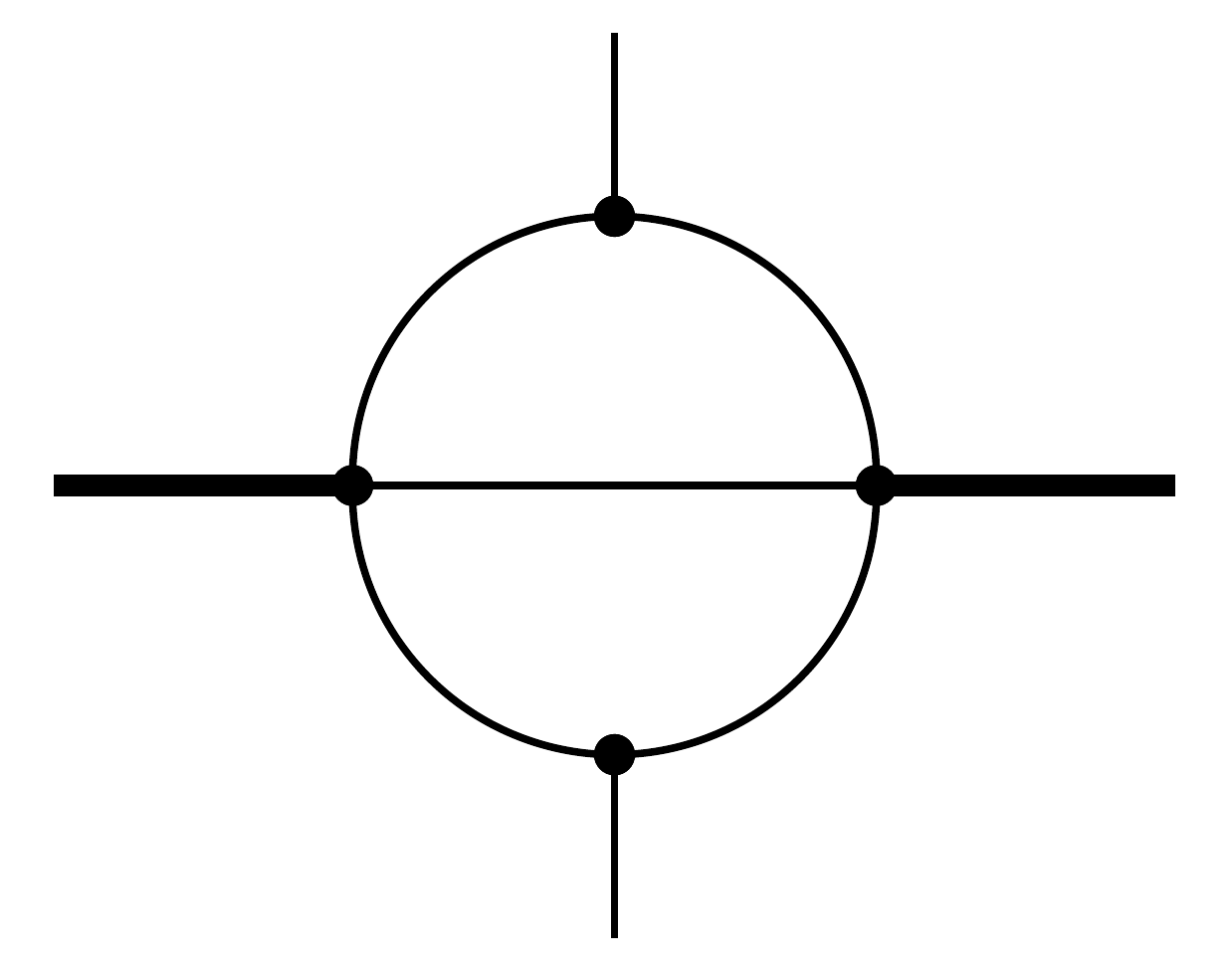} & \raisebox{2.5\height}{726} & \raisebox{2.5\height}{0.34}    \\
\hline
\includegraphics[width=0.1\textwidth]{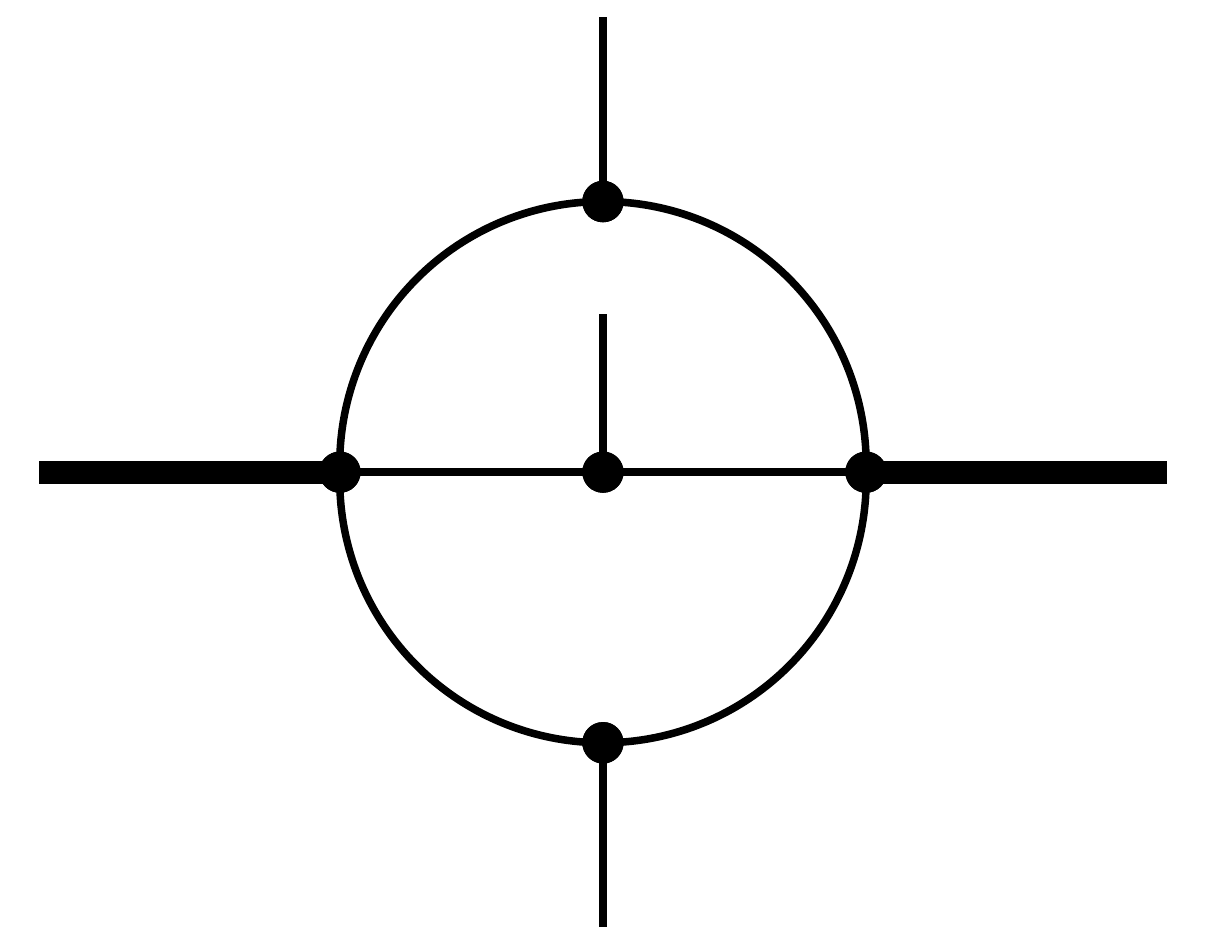} & \raisebox{2.5\height}{726} & \raisebox{2.5\height}{0.76}    \\
\hline
\includegraphics[width=0.1\textwidth]{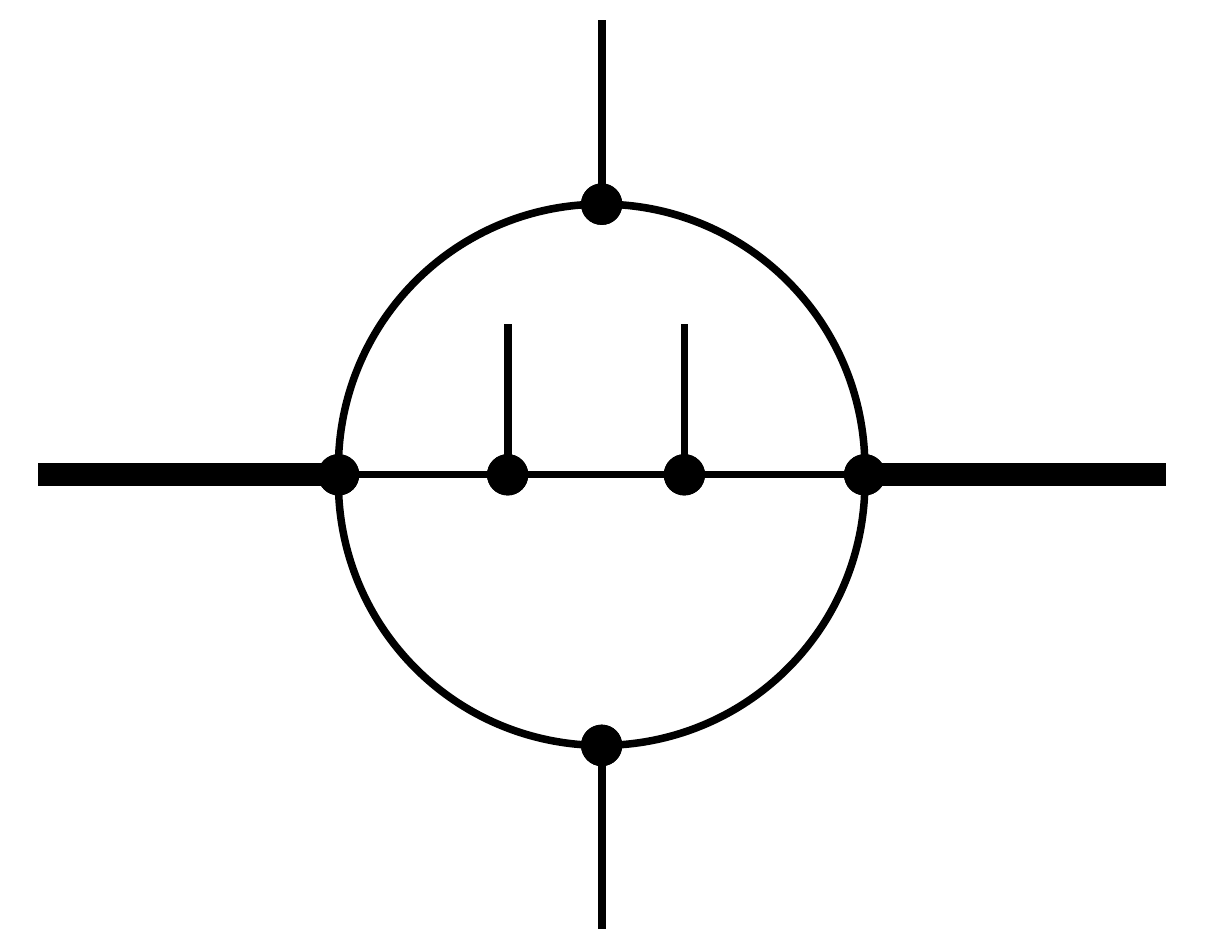} & \raisebox{2.5\height}{12826} & \raisebox{2.5\height}{2.91}    \\
\hline
\end{tabular}
\caption{Example of two-loop Feynman diagrams with external legs ranging from 2 to 6, where only the bold lines represent massive legs. The number of points required to achieve 6-digit precision and the average evaluation time (in millisecond) per point in double precision using dimension-changing transform \cite{DCT} are shown. }
\label{table: two-loop}
\end{table}

\sect{Two-loop examples}
To demonstrate the validity of our strategy, we apply it to two-loop FIs. In this case, Eq.~\eqref{eq:intMtilde} becomes
\begin{align}
\mathcal{M} = \int_0^1 \mathrm{d}X_1 \, \mathrm{d}X_2 \, \left[ \left( \tilde{\mathcal{M}} - \tilde{\mathcal{M}}_{\text{sub}} \right) + \tilde{\mathcal{M}}_{\text{sub}} \right],
\end{align}
where the subtraction term is given by
\begin{equation}
    \begin{aligned}
        \tilde{\mathcal{M}}&_{\text{sub}}=\sum_{\mu(\epsilon),n_\mu}X_1^{\mu(\epsilon)}\log(X_1)^{n_\mu}\sum_{i=0}^{\lfloor -\mu(0)\rfloor}X_1^i\tilde{M}_1^{\mu,n_\mu,i}(X_2)\\
        &+\sum_{\rho(\epsilon),n_\rho}X_2^{\rho(\epsilon)}\log(X_2)^{n_\rho}\sum_{j=0}^{\lfloor -\rho(0)\rfloor}X_2^j\tilde{M}_2^{\rho,n_\rho,j}(X_1)\\
        &-\sum_{\mu(\epsilon),n_\mu,\rho(\epsilon),n_\rho}X_1^{\mu(\epsilon)}\log(X_1)^{n_\mu}X_2^{\rho(\epsilon)}\log(X_2)^{n_\rho}\\
        & \times\sum_{i=0,j=0}^{\lfloor -\mu(0)\rfloor,\lfloor -\rho(0)\rfloor}X_1^i X_2^j \tilde{M}_3^{\mu,n_\mu,i,\rho,n_\rho,j}.
    \end{aligned}
\end{equation}
Here, $\mu(\epsilon)$ and $\rho(\epsilon)$ represent the singular behaviors of $X_1$ and $X_2$ that depend on $\epsilon$, with $\mu(0)$ and $\rho(0)$ denoting the corresponding values when $\epsilon = 0$. With this subtraction, \( \tilde{\mathcal{M}} - \tilde{\mathcal{M}}_{\text{sub}} \) is finite  throughout the entire integration region. The coefficients in \( \tilde{\mathcal{M}}_{\text{sub}} \) are determined by the differential equations of FBIs with respect to \( X_b \). The finite integrals are then evaluated using the adaptive Gaussian-Kronrod rule.

As an example, we have computed FIs defined by Tab.~\ref{table: two-loop} using our new method, ranging from two external legs to six external legs. We employ the adaptive Gaussian-Kronrod rule with degree 5 to achieve 6-digit precision. The number of points to be evaluated, as well as the computation time per point (using a single CPU), are provided in Tab.~\ref{table: two-loop}. Our results agree with that obtained using {\tt AMFlow} \cite{Liu:2022chg}. We note that this is a very preliminary implementation of the strategy, and significant improvements can be made in the future.

\sect{Summary and outlook}
In this Letter, we reveal a deep structure of Feynman integrals (FIs) by expressing them as integrals over a few branch parameters with a simple integrand. We have shown that the integrand, expressed as fixed-branch integrals (FBIs), is as straightforward as one-loop FIs, which can be computed directly. We also present a strategy to perform the remaining integration over the branch parameters. As a proof of concept, we successfully computed up to two-loop non-planar Feynman integrals with six external legs, demonstrating high efficiency.

It is important to emphasize that the application demonstrated in this Letter is only one of many possible uses for this novel representation of FIs. In fact, all existing methods for computing FIs can be applied to this representation, including integration-by-parts reduction ~\cite{Chetyrkin:1981qh,Tkachov:1981wb,Laporta:2000dsw,Anastasiou:2004vj,Smirnov:2008iw,Smirnov:2023yhb,Studerus:2009ye,Lee:2013mka,Maierhofer:2017gsa, Peraro:2019svx, Klappert:2020nbg,Wu:2023upw,Guan:2024byi}, asymptotic expansion \cite{Baikov:2005nv,Liu:2018dmc}, intersection theory \cite{Mizera:2017rqa,Mastrolia:2018uzb,Frellesvig:2019kgj,Frellesvig:2019uqt, Mizera:2019vvs, Mizera:2019ose, Frellesvig:2020qot, Caron-Huot:2021xqj, Caron-Huot:2021iev, Chestnov:2022alh, Fontana:2023amt,Brunello:2023rpq,Lu:2024dsb}, canonical differential equations \cite{Kotikov:1990kg,Remiddi:1997ny,Gehrmann:1999as,Henn:2013pwa}, auxiliary mass flow \cite{Liu:2017jxz,Liu:2021wks,Liu:2022mfb,Liu:2022chg}, and others. With these future developments, there is optimism that the challenges of computing multi-leg FIs beyond one-loop level can be overcome, enabling the high-precision requirements of particle physics experiments to be met.

\begin{acknowledgments}
	We would like to thank B. Feng, X. Guan and L.L. Yang for fruitful discussion. The work was supported by the National Natural Science Foundation of	China (No. 12325503), the National Key Research and Development Program of China under	Contracts No. 2020YFA0406400, the computing facilities at Chinese National Supercomputer Center in Tianjin and the High-performance Computing Platform of Peking University.
\end{acknowledgments}

\providecommand{\href}[2]{#2}\begingroup\raggedright\endgroup

\appendix

\section{Supplementary Material }

\sect{Derivation of IBP for FBIs}
When dealing with an integral that has a delta function within its measure and applying the integration by parts (IBP) method to this integral, we obtain the following expression:
\begin{equation}
    \begin{aligned}
        0=&\int_0^{\infty} \prod_{i=1}^n\mathrm{d}y_i\left(-\delta^{'}(1-\sum_{j=1}^n y_j)\;g(y_1,\cdots,y_n)+ \right.\\
        &\left.\delta(1-\sum_{j=1}^n y_j)\left(\frac{\mathrm{\partial}}{\mathrm{\partial}y_k}+\delta(y_k)\right)g(y_1,\cdots,y_n)\right).
    \end{aligned}
    \label{eq:1delta}
\end{equation}
For different derivatives $\frac{\mathrm{\partial}}{\mathrm{\partial}y_k}$ and $\frac{\mathrm{\partial}}{\mathrm{\partial}y_m}$, the first term of eq \eqref{eq:1delta} remains the same. Hence, by subtracting the expressions of eq \eqref{eq:1delta} corresponding to different variables from one another, we can derive the IBP identity that involves delta functions. This identity is given by:
\begin{equation}
    0=\int \left[\mathrm{d}\mathbf{y}\right]\left(\frac{\mathrm{\partial}}{\mathrm{\partial}y_k}+\delta(y_k)-\frac{\mathrm{\partial}}{\mathrm{\partial}y_m}-\delta(y_m)\right)g(\vec{y}).
    \label{eq:basicibp}
\end{equation}
In the case of FBIs, the variable $y_{(b,i)}$ belongs solely to branch b. As a result, when focusing on a particular branch, the FBI is still effectively equivalent to having only one delta function. Consequently, the IBP identity for FBIs take the following form: 
\begin{equation}
\begin{aligned}
    0& =-I_{\vec{\nu}-\vec{e}_{(b,i)}}^{\Delta-1}+I_{\vec{\nu}-\vec{e}_{(b,j)}}^{\Delta-1}+\sum_{\alpha=1}^N R_{(b,i),\alpha}\;\nu_\alpha I_{\vec{\nu}+\vec{e}_{\alpha}}^{\Delta}-\\
    &\;\sum_{\alpha=1}^N R_{(b,j),\alpha}\;\nu_\alpha I_{\vec{\nu}+\vec{e}_{\alpha}}^{\Delta}.
\end{aligned}
\label{eq:fbiIBP}
\end{equation}
The total number of independent IBP identities is $N-B$. However, this quantity is insufficient to reduce all the component of  $\vec{\nu}$ to lower indices. Additionally, due to the constraints of $\sum_i y_{(b,i)}=1$, we have the following identities for each branch:
\begin{equation}
    I_{\vec{\nu}}^{\Delta-1}=-\sum_{i=1}^{n_b}\nu_{(b,i)}\;I_{\vec{\nu}+\vec{e}_{(b,i)}}^{\Delta}.
    \label{eq:fbiscale}
\end{equation}
By combining and symmetrizing eq \eqref{eq:fbiIBP} and eq \eqref{eq:fbiscale}, we obtain eq \eqref{eq:rec}.

We introduce arbitrary constants $z_\alpha$ in order to satisfy the following system of linear equations:
\begin{equation}
    \sum_{\alpha=1}^N R_{(b,i),\alpha} \; z_\alpha=-C_b,\quad i=1,\cdots, n_b,\; b=1,\cdots, B,
\end{equation}
where $C_b$ is an arbitrary constant. We denote the sum of $z_{(b,i)}$ as $g_b=\sum_{i=1}^{n_b}z_{(b,i)}$. We set $j=1$ for all branch b in eq \eqref{eq:fbiIBP}, and multiply each term of equation \eqref{eq:fbiIBP} by $z_{(b,i)}$, and subsequently sum over all $(b,i)$, we arrive at the following relation:
\begin{equation}
    \begin{aligned}
        0=&-\sum_{\alpha=1}^{N} z_\alpha I_{\vec{\nu}-\vec{e}_\alpha}^{\Delta-1}+\sum_{b=1}^{B}g_b I_{\vec{\nu}-\vec{e}_{(b,j)}}^{\Delta-1}+C I_{\vec{\nu}}^{\Delta-1}\\
        &-\sum_{b=1}^{B}g_b \sum_{\alpha=1}^{N}R_{(b,j),\alpha}\;\nu_\alpha I_{\vec{\nu}+\vec{e}_\alpha}^{\Delta}.
    \end{aligned}
    \label{eq:sumIBP}
\end{equation}
If we change the seed of IBP to $y_{(b,i)}\prod_{\alpha=1}^{N}y_{\alpha}^{\nu_{\alpha}-1}\left(\mathcal{F}-\mathrm{i}0^+\right)^{\Delta-\nu}$ for branch b, and then sum over all $(b,i)$ except $(b,1)$ of all branches, we obtain:
\begin{equation}
    \begin{aligned}
        0&=\left(2\Delta-\nu-B\right)I_{\vec{\nu}}^{\Delta}+\sum_{b=1}^{B}I_{\vec{\nu}-\vec{e}_{(b,1)}}^{\Delta-1}-\\
        &\quad\sum_{b=1}^{B}\sum_{\alpha=1}^{N}R_{(b,1),\alpha}\;\nu_\alpha I_{\vec{\nu}+\vec{e}_{\alpha}}^{\Delta}.
    \end{aligned}
    \label{eq:additionibp}
\end{equation}
When we set $g_b=z_0$ for all branches, we then have the dimension shift relations for FBIs as given in eq \eqref{eq:dshift}.

\end{document}